\documentclass[runningheads]{llncs}
\usepackage[T1]{fontenc}
\usepackage{graphicx}
\usepackage{amsmath}

\usepackage[colorlinks=true, citecolor=blue, linkcolor=blue, urlcolor=blue]{hyperref}
\usepackage{color}

\usepackage{xcolor}

\usepackage[absolute,overlay]{textpos}
\usepackage{xcolor}

\begin{document}
\begin{textblock*}{210mm}(0mm,12mm) 
    \centering
    \begin{minipage}{13cm} 
        \centering
        \textcolor{red}{\large \textbf{The work is accepted for publication as a full paper (Main Track) at the 27th International Conference on Artificial Intelligence in Education (AIED 2026).}}
    \end{minipage}
\end{textblock*}
\title{Automatically Inferring Teachers' Geometric Content Knowledge: A Skills Based Approach}
\titlerunning{Automatically Inferring Teachers' Geometric Content Knowledge}
\author{Ziv Fenigstein\inst{1}\orcidID{0009-0002-2237-1120} \and
Kobi Gal\inst{1,2}\orcidID{0000-0001-7187-8572} \and
Avi Segal\inst{1}\orcidID{0000-0003-1422-2598} \and
Osama Swidan\inst{1}\orcidID{0000-0002-2689-7173} \and
Inbal Israel\inst{1}\orcidID{0009-0005-3213-8922} \and
Hassan Ayoob\inst{1}\orcidID{0009-0005-7726-6602}}

\authorrunning{Z. Fenigstein et al.}

\institute{Ben-Gurion University, Israel\\
\email{zivfenig@post.bgu.ac.il, kobig@bgu.ac.il, avise@post.bgu.ac.il, osamas@bgu.ac.il, inbalalmasy@gmail.com, Hassan.ayoob@gmail.com} \and
University of Edinburgh, U.K.\\
\email{kgal@ed.ac.uk}}

\maketitle              
\begin{abstract}
Assessing teachers' geometric content knowledge is essential for geometry instructional quality and student learning, but difficult to scale. The Van Hiele model characterizes geometric reasoning through five hierarchical levels. Traditional Van Hiele assessment relies on manual expert analysis of open-ended responses. 
This process is time-consuming, costly, and prevents large-scale evaluation. This study develops an automated approach for diagnosing teachers' Van Hiele reasoning levels using large language models grounded in educational theory.
Our central hypothesis is that integrating explicit skills information significantly  improves Van Hiele classification. 
In collaboration with mathematics education researchers, we built a structured skills dictionary decomposing the Van Hiele levels into 33 fine-grained reasoning skills. 
Through a custom web platform, 31 pre-service teachers solved geometry problems, yielding 226 responses. Expert researchers then annotated each response with its Van Hiele level and demonstrated skills from the dictionary. Using this annotated dataset, we implemented two classification approaches: (1) retrieval-augmented generation (RAG) and (2) multi-task learning (MTL). Each approach compared a skills-aware variant incorporating the skills dictionary against a baseline without skills information. Results showed that for both methods, skills-aware variants significantly outperformed baselines across multiple evaluation metrics. 
This work provides the first automated approach for Van Hiele level classification from open-ended responses. 
It offers a scalable, theory-grounded method for assessing teachers' geometric reasoning 
that can enable large-scale evaluation 
and support adaptive, personalized teacher learning systems.

\end{abstract}

\section{Introduction}

Teachers’ content knowledge (CK) in mathematics has been shown to directly impact students' learning outcomes~\cite{campbell2011impact,copur2024impact,lumbre2023relationship}. 
Within the realm of Geometry,  pre-service teachers often demonstrate lower competency  than expected for effective instruction~\cite{armah2018investigating,kurt2020analysis,Manero2021}. Understanding teachers’ content knowledge in this domain is therefore critical for both research and professional development.

The Van Hiele model of geometric thought characterizes geometry content knowledge through five hierarchical levels of reasoning ranging from visualization to formal rigor~\cite{vanhiele1959}.
Building on this framework, researchers have argued that Van Hiele assessment should focus on analyzing learners' responses to open-ended questions rather than focus on the correctness of their answers~\cite{crowley1987van,jaime1994model}.
Although widely used in education research, existing approaches for assessing Van Hiele levels rely primarily on manual response analysis, which is time-consuming, costly, and hard to scale~\cite{kurt2020analysis,lumbre2023relationship,Manero2021}.

This paper addresses this gap by using Large Language Models (LLMs) to classify teachers' Van Hiele levels from their open-ended responses.
 The  central hypothesis of our work was that integrating explicit skills information  would significantly improve Van Hiele classification.
To this end, we constructed a structured skills dictionary in collaboration with mathematics education researchers, decomposing the five Van Hiele levels into a total of 33 fine-grained reasoning skills that characterize each level.
We first conducted a data collection study with 31 pre-service math teachers who provided open-ended responses to geometry problems through a custom web-based platform, yielding 226 question-response pairs.
Expert researchers in mathematics education annotated each pair with its Van Hiele level and identified which of that level's associated skills from the skills dictionary were demonstrated in the response.
We then developed several classification models that, given a question-response pair, diagnose the Van Hiele level reflected in the response rather than just surface correctness.

We considered several approaches for classification, (1) retrieval-augmented generation (RAG), which incorporates skills by retrieving annotated examples with skills labels and including the skills dictionary in the model's prompt, and (2) multi-task learning, which incorporates skills through attention mechanisms and an auxiliary skills prediction task alongside the primary Van Hiele level classification. For each classification approach, we compared a skills-aware variant against a baseline without skills information, isolating the contribution of explicit skills modeling to classification performance.
For both classification approaches, the skills-aware variants significantly outperformed the baseline variants on held out test-sets. 
To better understand the models' behavior,  we examined the models' sensitivity to skill definitions in the RAG approach, the contribution of individual skill components in the multi-task learning approach, and classification difficulty patterns across Van Hiele levels.

Our work makes several contributions. (1) it is the first automated approach for inferring Van Hiele levels from open-ended responses, addressing a scalability challenge in geometry education research; 
(2) it provides two novel classification approaches demonstrating that explicitly  modeling  teachers' skills significantly improves classification regardless of the modeling paradigm; and (3)  it provides new data sources for researchers in collaboration with mathematics education researchers: a  theoretically grounded skills dictionary and an annotated dataset of 226 question-response pairs from pre-service teachers. 
Our code, models, data, prompts, and skills dictionary are publicly available in GitHub\footnote{\url{https://github.com/zivfenig/Van-Hiele-Level-Classification}}.

\section{The Van Hiele Model of Geometric Reasoning}
\label{Van_Hiele_Theory}
The Van Hiele model~\cite{vanhiele1959} describes five hierarchical levels of geometric reasoning demonstrated by learners. 
Advancement through levels depends on mastery of the preceding levels. At the most basic level (Level~1, Visualization), learners identify shapes based on their overall appearance. For example, a student recognizes a rectangle as visually distinct from a trapezoid. At Level~2 (Analysis), learners recognize and describe properties of geometric shapes. For example, a student might state that a square has four right angles and all sides of equal length. Level~3 (Informal Deduction) involves reasoning about relationships between properties and between shape classes. For example, a student might conclude that a rectangle is a type of parallelogram because it possesses all the defining properties of parallelograms. At Level~4 (Deduction), learners construct formal proofs based on definitions, axioms, and theorems. For example, a student can prove that the opposite angles of a parallelogram are congruent. The highest level (Level~5, Rigor) entails abstract reasoning in non-Euclidean or axiomatic systems. For example, a student analyzes how geometric properties, concepts, or proofs change when axioms are modified or when comparing Euclidean and non-Euclidean geometries~\cite{Jupri_2018}.
Although the Van Hiele model was originally developed to characterize geometric reasoning in school-aged students, its levels reflect instructional experience rather than age or maturation~\cite{crowley1987van}. This makes it applicable to learners at any stage, including teachers. Mayberry~\cite{mayberry1983van} confirmed this empirically, showing that the Van Hiele hierarchy holds for undergraduate preservice teachers. Since teachers' geometric content knowledge directly shapes their students' learning opportunities~\cite{kurt2020analysis,beswick2012measuring}, understanding teachers' Van Hiele levels is essential for improving geometry instruction. The model has therefore been applied to teachers for a range of purposes: (1) inferring their content knowledge in geometry~\cite{kurt2020analysis,Manero2021}; (2) measuring the relationship between their geometric reasoning and student achievement~\cite{lumbre2023relationship}; and (3) studying the effects of pedagogical interventions on their content knowledge~\cite{armah2018investigating,swafford1997increased}, and examining their ability to diagnose students' geometric understanding~\cite{yi2020examining}.
All of these  studies have relied on either manual expert assessment of open-ended responses or multiple-choice questionnaires. Manual assessment is a time-consuming and resource-intensive process  that limits sample sizes and prevents large-scale longitudinal tracking of teacher development. Automating this classification enables researchers to scale up their studies significantly.

Usiskin~\cite{usiskin1982van} developed a multiple-choice test where learners at a given Van Hiele level were expected to correctly answer items at that level and all preceding levels. 
However, researchers have critiqued this approach, arguing that Van Hiele levels comprise multiple distinct reasoning processes~\cite{crowley1987van,gutierrez1991alternative}. Therefore, accurate assessment requires evaluating the specific skills demonstrated in learners' explanations, not merely correctness on multiple-choice items. This skills-based perspective directly motivates our modeling approach of decomposing each Van Hiele level into explicit, fine-grained reasoning skills.

For example, consider the question: \textit{``Given a shape with four equal sides, is it necessarily a rhombus?''} One student might respond, \textit{``No, maybe it is a square''}. This reflects Level~2 reasoning (Analysis) demonstrating the Level 2 skill of describing shapes in terms of their properties but not the Level 3 skill of recognizing that squares are a special case of rhombus. Another student might say, \textit{``Yes, because a square is also a type of rhombus''}, demonstrating Level~3 type reasoning (Informal Deduction) by explicitly understanding the inclusion relationship, that one shape class can be a subset of another. 

Building on this paradigm, our work develops a structured skills dictionary decomposing each Van Hiele level into explicit reasoning skills in collaboration with mathematics education researchers. This skill decomposition forms the foundation of our automated classification approach for answers to open-ended questions (the full skills dictionary available in our repository\footnote{\url{https://github.com/zivfenig/Van-Hiele-Level-Classification}}).

\section{Related Work}

Recent advances have made LLMs increasingly effective for educational assessment. Henkel et al.~\cite{henkel2025can} demonstrated that LLMs can achieve near-parity with human raters when grading open-ended reading comprehension responses, distinguishing between fully correct, partially correct, and incorrect responses. Beyond correctness, Lee et al.~\cite{lee2024applying} utilized LLMs to classify middle-school written responses to science assessments into proficiency levels (Proficient, Developing, and Beginning), finding that Chain-of-Thought prompting with explicit rubrics improved classification 
accuracy across proficiency levels.
However, LLMs can struggle with nuanced reasoning assessment. Rachmatullah et al.~\cite{rachmatullah2026exploring} used LLMs to assess teachers' Pedagogical Content Knowledge (PCK). They found that while  LLMs achieved high reliability when scoring lesson plans, they struggled to assess teachers' analyses of student misconceptions, relying heavily on keyword matching and therefore failed to accurately identify more nuanced responses from teachers which did not include those keywords.
This highlights the challenge of capturing nuanced reasoning. Our skills-aware approach addresses this by explicitly defining the reasoning patterns that distinguish between Van Hiele levels. While these studies demonstrate LLMs’ potential for educational assessment, to our knowledge no prior work has applied LLMs or machine learning to assess geometric reasoning under the Van Hiele framework.

RAG has proven effective for classifying open-ended student responses. Fateen et al.~\cite{fateen2024beyond} employed RAG to retrieve similar student answers as few-shot examples for automatic short-answer scoring, improving performance over baseline methods. Additionally, Jauhiainen and Guerra~\cite{jauhiainen2024evaluating} utilized RAG to retrieve relevant reference material from scholarly articles for grading university examination answers. We build on these approaches by augmenting retrieved examples with explicit skills annotations, providing pedagogically grounded context that guides classification beyond surface similarity.

Multi-task learning is a well-established and effective paradigm across AI domains. Xu et al.~\cite{xu2022multi} demonstrated that attention mechanisms informed by auxiliary task outputs improve performance by incorporating structured domain knowledge in image classification and segmentation tasks. This approach has been successfully applied to education research. Huang et al.~\cite{huang2020neural} showed that jointly learning multiple teacher-question classification tasks improves performance through shared semantic representations. An et al.~\cite{An_Kim_Kim_Park_2022} and Geden et al. ~\cite{geden2020predictive} demonstrated that auxiliary objectives - such as option tracing or per-item prediction, lead to more accurate and stable student assessment.
 Inspired by these works, we apply multi-task learning where skills serve as an auxiliary prediction task alongside Van Hiele level classification.

\section{Methodology}
\begin{figure}
    \centering
    \includegraphics[width=\textwidth]{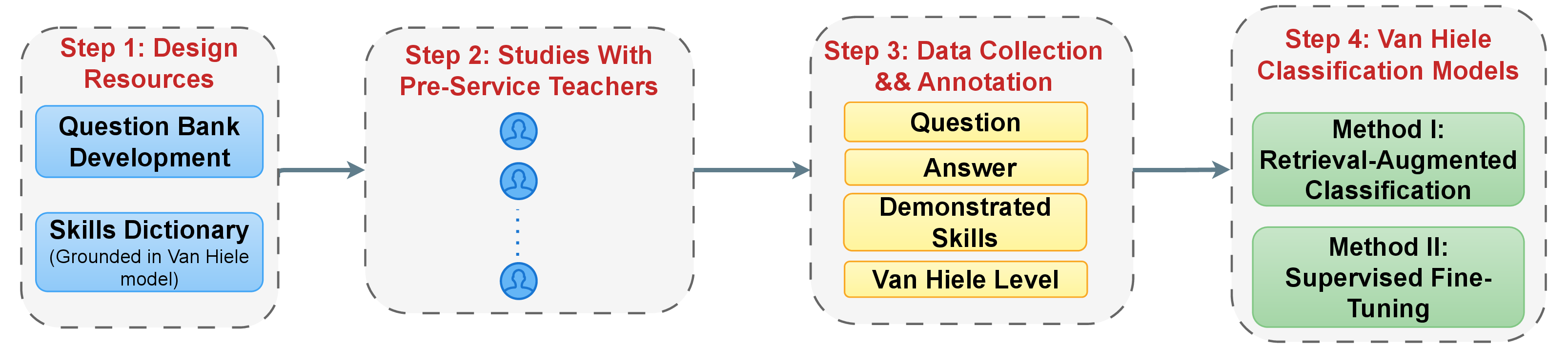}
    \caption{Overview of the proposed skills-based Van Hiele classification framework.}
    \label{fig:methodology_overview}
\end{figure}

In this section, we present our methodology to automatically classify teachers' Van Hiele levels using skills-aware modeling approaches (see Figure~\ref{fig:methodology_overview}).
We first developed a comprehensive geometry question bank and structured skills dictionary grounded in Van Hiele theory. We then collected and annotated a dataset of teacher responses with expert-assigned Van Hiele levels and demonstrated skills.  
Using this annotated dataset we developed the two classification methods - one using retrieval-augmented generation, the other using multi-task learning - each compared against a baseline without skills information to isolate the contribution of explicit skills modeling.  We studied whether the skills-aware variant of both methods outperformed the baselines.

\subsection{Question Bank and Skills Dictionary Development}
\label{sec:question_bank}
We developed a question  bank building on problems from Usiskin’s Van Hiele geometry test, which consists entirely of closed-ended multiple-choice items~\cite{usiskin1982van}. We adapted items into open-ended formats to elicit richer explanations from teachers, promoting reasoning beyond recognition of fixed options.

In total we constructed 59 different open-ended geometry problems.  
The questions addressed widely taught geometry topics such as quadrilaterals, angle relationships, triangle congruence, and similarity. 
Items varied in difficulty to elicit reasoning across the Van Hiele spectrum. 
Example of question included \emph{``List at least two properties that are shared by all squares but not shared by all rhombus, justify your choices''}.

We built a structured skills dictionary in collaboration with three mathematics education 
researchers with expertise in geometry education and experience teaching and studying 
pre-service mathematics teachers. The skills dictionary was grounded in Crowley's 
theoretical decomposition of Van Hiele levels~\cite{crowley1987van}, which establishes that 
each level is characterized by specific observable reasoning behaviors and distinct linguistic 
markers that can only be reliably identified from open-ended explanations. For example, 
Crowley explicitly identifies level-specific vocabulary and behaviors that directly correspond 
to skills in our dictionary, such as the use of logical connectives (``if\ldots then'', 
``it follows that'') at the Informal Deduction level, and the ability to identify what is 
given versus what must be proved at the Deduction level. Drawing on this theoretical foundation and their pedagogical experience with pre-service teachers, the experts decomposed each Van Hiele level into fine-grained reasoning skills that characterize that level, resulting in 33 distinct skills across the five levels.

To collect teachers' responses, we developed a custom web-based platform. The interface presents a sequence of problems from our question bank and teachers provide responses in free text, allowing them to articulate their full reasoning process.
For problems requiring geometric construction or manipulation, the platform embeds an interactive GeoGebra~\cite{Tamam_2021geogebra} applet alongside the question.
The applet served only as a cognitive scaffold and was not used in our model.

\subsection{Data Collection and  Annotation}
A total of 31 pre-service mathematics teachers from three regional teacher training institutions participated in our study. All participants were enrolled in a geometry course. The study protocol was reviewed and approved by the Institutional Review Board (IRB). All participants provided informed consent before taking part in the study. 
The web-based platform described in Section~\ref{sec:question_bank} served both for data collection and geometry practice, incorporating multiple-choice (MCQ) and open-ended problems. Each participant was assigned 20 problems randomly selected (at least 10 open-ended) from our question bank, including problems of varying difficulty designed to elicit responses across all Van Hiele levels. This ensured coverage of the full spectrum of geometric reasoning, from basic visualization (Level~1) to Rigor (Level~5). The platform supported flexible participation with no strict time limits and the ability to save and resume work. For this study, we retained only open-ended responses; MCQ responses and empty submissions were excluded, yielding 226 valid question-response pairs.

We employed a double-blind annotation protocol to ensure objective ground-truth labeling of Van Hiele levels. Two experts in mathematics education independently reviewed every response and labeled it with the Van Hiele level without seeing the other's labels. 
The inter-rater reliability was high (Cohen’s $\kappa = 0.84$) indicating substantial agreement between annotators' initial labels.
After assigning the Van Hiele level, the two experts collaboratively identified which skills from the skills dictionary were demonstrated in each response. 


This annotation process produced a dataset of 226 question-response pairs, each annotated with a Van Hiele level and demonstrated skills. The dataset contains responses across all five levels, with Levels 2 and 3 most common and Levels 4 and 5 less represented, aligning with past studies~\cite{armah2018investigating,Manero2021}.


\subsection{Van Hiele Level Classification Models}

We developed two methods to classify Van Hiele levels from question–response pairs. For each, we implemented a baseline without skills information and a skills-aware variant incorporating our structured skills dictionary. All other components are identical across variants, isolating the effect of skills information.

\subsubsection{Method I: Retrieval-Augmented Classification.}
\begin{figure}
    \centering
    \includegraphics[width=\textwidth]{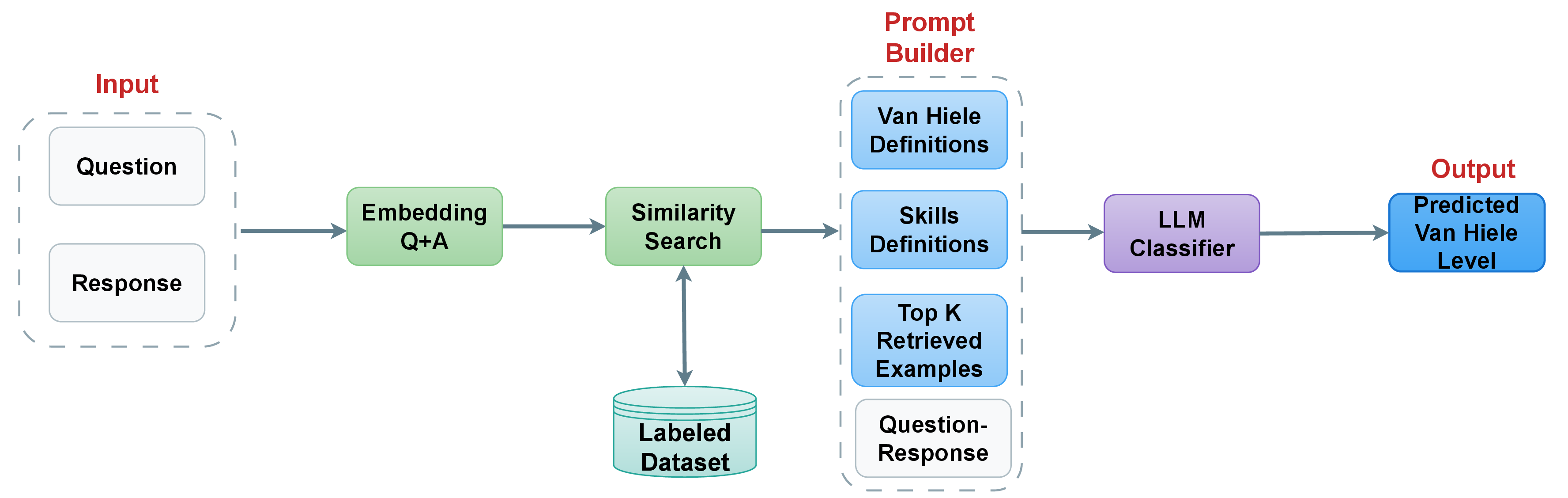}
    \caption{Retrieval-augmented generation architecture for Method I. }
    \label{fig:method1_rag_pipeline}
\end{figure}

The first method uses a Retrieval-Augmented Generation (RAG) pipeline that receives a new question-response pair as input, retrieves similar pairs from our annotated dataset, and integrates them into a prompt for LLM classification. Figure~\ref{fig:method1_rag_pipeline} provides an overview of this retrieval-augmented classification architecture.
Each question-response pair in the dataset is encoded separately using  \texttt{multilingual-e5-base}\footnote{\url{https://huggingface.co/intfloat/multilingual-e5-base}} embedding model. All embeddings are $L_2$-normalized to optimize retrieval within the RAG pipeline, enabling efficient and stable cosine similarity computation via normalized dot products~\cite{faisslibrary}.  

For a new question-response pair, we construct a weighted query embedding assigning 80\% weight to the response and 20\% to the question. This weighting reflects the Van Hiele framework's emphasis on reasoning expressed in responses rather than questions themselves (see discussion in Section \ref{Van_Hiele_Theory}). Cosine similarity is computed between the new question-response embeddings and the stored vectors, and the top-K (see Section~\ref{sec:settings} for K) most similar examples are retrieved.

In the \emph{baseline variant} (RAG without skills), each retrieved example includes the geometry question, the teacher’s response, and the corresponding Van Hiele level. The LLM is also supplied with a  system prompt containing only the standard Van Hiele level definitions. No skill annotations or skill definitions are provided to the model. 
In the \emph{skills-aware variant}, retrieved examples additionally contain the annotated skills associated with each teacher's response, and the system prompt includes both Van Hiele level definitions and the full skills dictionary.

All other components of the pipeline - including the input encoding, retrieval procedure, number of retrieved examples, language model, and inference settings - are the same for both variants.  As a result, any observed performance differences between the baseline and skills-aware versions can be attributed specifically to the presence of explicit skills information in the prompt.

\subsubsection{Method II: Multi-Task Learning.}
\begin{figure}
    \centering
    \includegraphics[width=\textwidth]{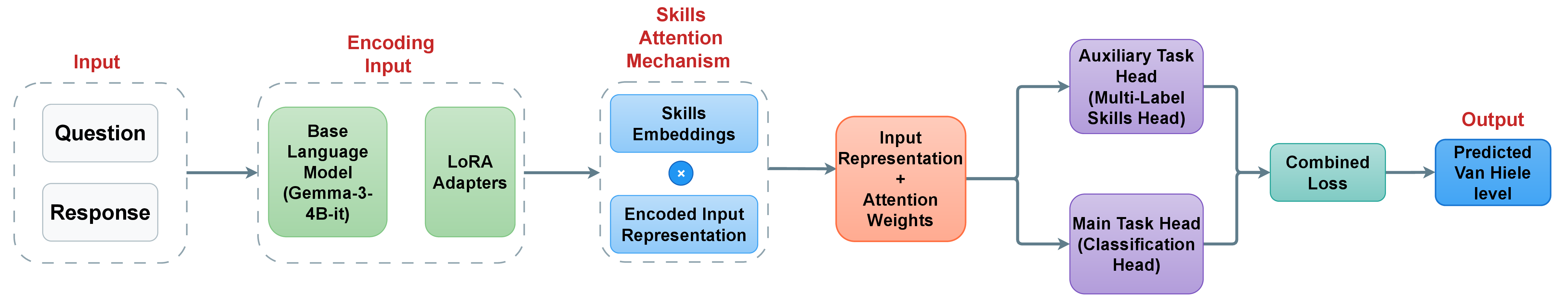}
    \caption{Multi-task learning architecture for Method II. 
    }
    \label{fig:method2_multitask_pipeline}
\end{figure}

The second classification method (see Figure \ref{fig:method2_multitask_pipeline}) fine-tunes an open-source LLM for Van Hiele classification. 
Similarly to Method I, the model receives a question-response pair as input and outputs the Van Hiele level.
We implemented two variants: a baseline that performs classification without skills information during training, and a skills-aware variant that incorporates an auxiliary skills prediction task with Van Hiele classification. 
Both variants used the same base open-source LLM.

In the \emph{baseline variant}, the question-response pair is encoded by the LLM and passed directly to a linear classification head that predicted the Van Hiele level. This variant does not use skills information during training. 

In the \emph{skills-aware variant},  we augmented the baseline with a skills attention mechanism and an auxiliary skills prediction task. Like the baseline, the question-response pair is first encoded by the LLM to produce an input representation.
Each Van Hiele skill from the skills dictionary is represented by a trainable embedding vector, initialized from text encodings of the skill descriptions using the \texttt{multilingual-e5-base} model. Attention weights are computed via dot products between the LLM's encoded input representation and the skill embeddings followed by a softmax normalization to produce a probability distribution over skills, quantifying how strongly the question-response pair aligns with each skill. 
This attention weights vector is then concatenated with the input representation (encoded question-response pair). The enriched representation feeds two prediction heads in parallel: a primary Van Hiele level classification head (identical to the baseline) and an auxiliary skills prediction head that predicts which skills are demonstrated.
The model is trained by optimizing a combined loss balancing the Van Hiele classification and auxiliary skills prediction tasks, weighted by  $\lambda$:
\begin{equation}
    L_{\text{total}} =  L_{\text{level}} + \lambda \cdot L_{\text{skills}}
\end{equation}
where $L_{\text{level}}$ is cross-entropy loss for Van Hiele level prediction, penalizing incorrect classification across the five distinct levels, and $L_{\text{skills}}$ is binary cross-entropy loss for skills prediction, penalizing incorrect presence/absence predictions for each skill independently. 
See Section~\ref{sec:settings} for implementation details.

These components create complementary learning signals during training. 
The attention mechanism identifies which skills matter for each input, while the auxiliary prediction task ensures the model learns representations that encode these skills. 
The dual learning signals from the classification loss and the skills prediction loss shape both the LLM's representations and the skills embeddings through backpropagation to better capture Van Hiele-specific reasoning patterns.

At inference, both variants take the question-response pair as input and output the predicted Van Hiele level. The skills-aware variant also computes skills predictions and attention weights internally, which can be accessed for interpretability analysis if needed. 
As in Method I, all other components and hyperparameters were the same for both variants.

\subsection{Implementation Settings}
\label{sec:settings}

For Method I, all retrieval-augmented variants use \texttt{Gemini-2.0-Flash}~\cite{team2023gemini}, chosen for long-context processing. Retrieval uses $K=5$ examples, as prior work shows this optimizes citation recall~\cite{leto2024toward}. Generation uses temperature~0.0 for deterministic predictions~\cite{bajan2025exploring}. Variants differ only in prompt content: the baseline includes Van Hiele level definitions and retrieved examples annotated with levels, while the skills-aware variant additionally includes the skills dictionary and skill annotations. Gemini 2.0 Flash was accessed via Vertex AI on Google Cloud Platform.

For Method II, both fine-tuning variants use \texttt{Gemma-3-4B-IT}~\cite{team2025gemma}, a mid-sized instruction-tuned model balancing capacity and efficiency for limited-data supervision. We apply LoRA (rank~16), which performed best empirically, keeping base weights frozen.

To mitigate overfitting risk given the limited dataset size, we apply dropout in LoRA adapters (0.05) and in the Van~Hiele classification head (0.25), with learning rate $2 \times 10^{-4}$ and weight decay~0.05. Training runs up to 30 epochs with early stopping on macro F1 (patience=4). The skills-aware variant uses auxiliary loss weight $\lambda=0.5$, selected empirically. Both variants were trained on an NVIDIA~RTX~6000 GPU.
Additional technical configurations can be found in our GitHub
repository\footnote{\url{https://github.com/zivfenig/Van-Hiele-Level-Classification}}.

\section{Experiments And Results}
 
We evaluated both methods by comparing their baseline and skills-aware variants on classifying Van Hiele levels in our collected dataset. We used five-fold cross-validation with a fixed random seed (42) for reproducibility. The dataset was partitioned into five independent folds while preserving the Van Hiele level distribution. Each fold served as a test set in turn, with the remaining four folds used for training and retrieval (for the RAG system). Within each training set, 15\% was held out as a validation set for parameter tuning and early stopping.

We report macro-averaged and weighted-averaged F1 scores (F1-macro, F1-weighted) to assess standard classification performance. F1-macro measures how consistently the model performs across Van~Hiele levels by assigning equal weight to each class, while F1-weighted reflects overall performance by accounting for the empirical label distribution. To capture the ordinal structure of the Van~Hiele hierarchy, we additionally report 
Mean Absolute Error (MAE), penalizing larger ordinal misclassifications, and 
Quadratic Weighted Kappa (QWK), which accounts for chance agreement; QWK ranges 
from $-1$ to $1$, where values close to $1$ indicate stronger agreement. 

\begin{figure}
    \centering
    \includegraphics[width=\textwidth]{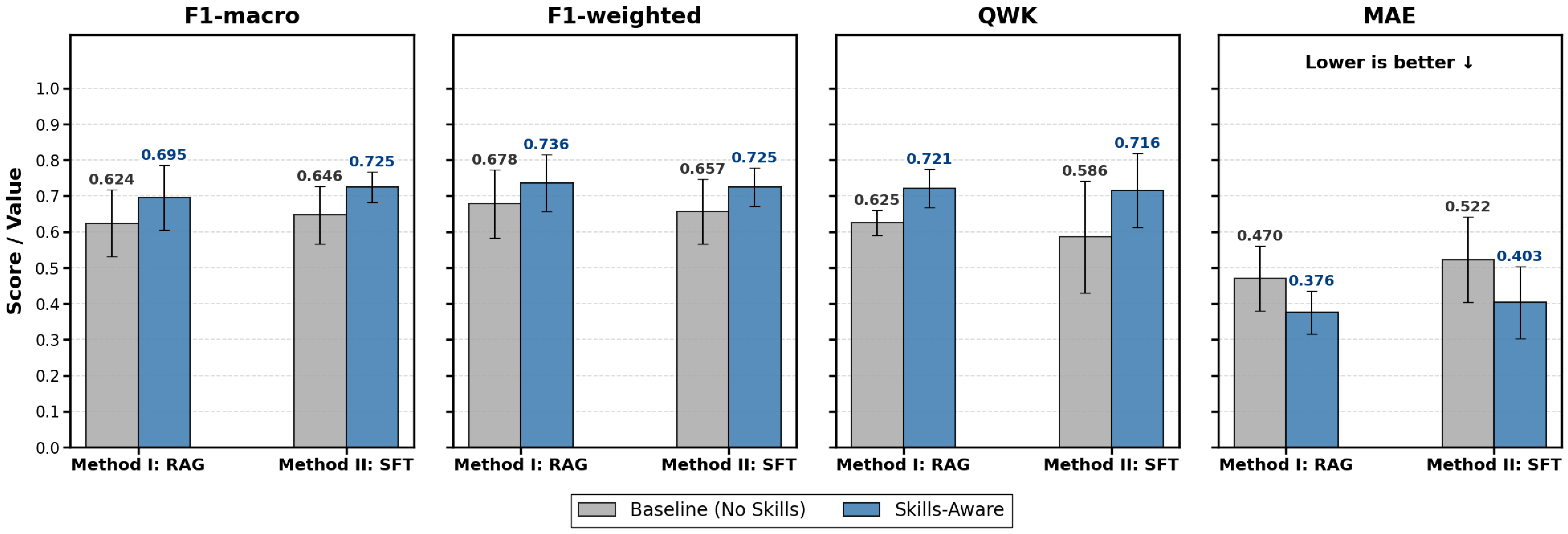}
    \caption{Average results across 5-fold cross-validation comparing baseline and skills-aware variants for RAG (Method I) and MTL (Method II). Numbers above bars show mean scores; lines within the bars indicate standard deviation. For MAE measure, lower is better.}
    \label{fig:main_results_plots}
\end{figure}

Figure \ref{fig:main_results_plots} shows the average results for RAG classification (Method I) and Multi-task learning methods (Method II), comparing baseline (gray bars) against skills-aware (blue bars) classification across five-fold test sets.  
For both methods, the  skills-aware models significantly outperformed the relevant baseline in all measures. 
We also note that the standard deviations of the  skills-aware variants in both methods  were consistently lower compared to their respective baselines, suggesting the models are more stable than the baseline variants.

A series of paired t-tests confirmed that the improvements of Method I are statistically significant across all metrics: F1-macro (t(4)=2.909, p=0.0437), F1-weighted (t(4)=3.008, p=0.0396), MAE (t(4)=-5.2, p=0.0065), and QWK (t(4)=3.465, p=0.0257).  
With respect to Method II, statistical significance was obtained  for F1-macro (t(4)=2.908, p=0.043), MAE (t(4)=-3.836, p=0.0185) and  QWK (t(4)=3.874, p=0.018), but not for F1-weighted measure. 
A possible reason for this is the small data set (226 question-response pairs).


 \subsection{Sensitivity Analysis}
 We performed several sensitivity analyses to study how variations in skill-related information influence model behavior.
With respect to Method I, we wanted to validate  the performance gain of the skills-aware variant to selecting the \emph{right} skills for the \emph{right} question-response pair, as opposed to simply providing the model with additional context. To this end, we compared the baseline and the skills-aware variants to a  ``noisy skills'' variant. In this setup, we randomly shuffled the skills definitions,  so that each skill label was assigned a definition of another skill. All other components of the skill-aware method and the empirical methodology remained the same. 
\begin{table}
\centering
\caption{Skills sensitivity analysis for Method I. Values are reported as mean $\pm$ standard deviation across five cross-validation folds.}
\label{tab:method1_corrupted_skills}
\begin{tabular}{|l|c|c|c|c|}
\hline
\textbf{Variant} & \textbf{F1-macro} & \textbf{F1-weighted}  & \textbf{QWK} & \textbf{MAE} \\
\hline
Baseline (No Skills) 
& $0.624 \pm 0.093$ 
& $0.678 \pm 0.095$   
& $0.625 \pm 0.035$
& $0.47 \pm 0.09$\\
\textbf{Skills-Aware}      
& $\mathbf{0.695 \pm 0.09}$ 
& $\mathbf{0.736 \pm 0.08}$ 
& $\mathbf{0.721 \pm 0.054}$
& $\mathbf{0.376 \pm 0.06}$ \\
Noisy Skills
& $0.61 \pm 0.122$
& $0.635 \pm 0.117$
& $0.673 \pm 0.093$
& $0.495 \pm 0.156$\\
\hline
\end{tabular}
\end{table}
Table~\ref{tab:method1_corrupted_skills} shows that the noisy skills variant exhibited significant performance degradation across all metrics compared to the skills-aware, and even performed worse than baseline. This demonstrates that performance gains depend on the model’s ability to utilize correctly aligned and pedagogically meaningful skills information, rather than simply benefiting from additional contextual input or increased prompt length.

For Method II, we quantified the individual contributions of the skills attention mechanism and auxiliary skills prediction head by isolating each component. We compared two partial variants: (1) \emph{Attention-Guided} variant,  trained with the skills-based attention mechanism and the Van Hiele classification head, without the auxiliary skills prediction task; (2) \emph{Skills-Supervised} variant, trained with the auxiliary skills prediction head and the Van Hiele classification head,  but without the skills-based attention mechanism. Table \ref{tab:method2_ablation} compares the full approach \emph{Full Model (Method II)} (skills-aware variant) to the \emph{Attention-Guided} , the \emph{Skills-Supervised} variants and the \emph{Baseline (No Skills)} variant. 
 \begin{table}
\centering
\caption{Skills component analysis for Method~II. Values are reported as mean $\pm$ standard deviation across five cross-validation folds. }

\label{tab:method2_ablation}
\begin{tabular}{|l|c|c|c|c|}
\hline
\textbf{Variant} 
& \textbf{F1-macro} 
& \textbf{F1-weighted} 
& \textbf{QWK}
& \textbf{MAE}\\
\hline
Baseline 
& $0.646 \pm 0.082$ 
& $0.657 \pm 0.092$ 
& $0.586 \pm 0.157$
& $0.523 \pm 0.119$ \\

Attention-Guided 
& $0.676 \pm 0.058$ 
& $0.679 \pm 0.066$  
& $0.632 \pm 0.129$
& $0.487 \pm 0.086$\\

Skills-Supervised 
& $0.652 \pm 0.063$ 
& $0.653 \pm 0.088$ 
& $0.581 \pm 0.089$
& $0.558 \pm 0.119$ \\

\textbf{Full Model (Method II)} 
& $\mathbf{0.725 \pm 0.043}$ 
& $\mathbf{0.725 \pm 0.053}$ 
& $\mathbf{0.717 \pm 0.103}$ 
& $\mathbf{0.403 \pm 0.102}$ \\
\hline
\end{tabular}
\end{table}

Results show the attention-guided variant clearly outperforms the baseline, while skills supervision alone yields limited gains. The full model performs best, indicating the components are complementary: the auxiliary task encourages the encoder to capture patterns aligned with pedagogical skills and Van Hiele levels, and the attention mechanism leverages these signals to focus on skills-relevant aspects of responses, producing more informative representations for level classification.
 
\subsection{Error Analysis}
To identify which Van Hiele levels are most difficult to classify, we analyzed per-level performance by aggregating predictions across all cross-validation folds. 
Both methods achieve high accuracy on Level~4. A possible reason is that Level~4 
(Deduction) responses involve constructing formal proofs, which are characterized by 
structured patterns that are easier for models to learn and identify. The skills dictionary 
likely contributes to this: Level~4 skills such as identifying given information versus what must be proved, and constructing formal logical arguments, map directly onto textual patterns that models can detect reliably.

In addition, for both methods, Level 2 accuracy is around 60\%, with most misclassifications occurring as Level 3. This demonstrates the difficulty of distinguishing between these adjacent levels, where differences in responses may be subtle and evident only in specific linguistic details (see example in Section \ref{Van_Hiele_Theory}).
 
Level 5 is challenging for both methods but shows markedly different performance: Method~I 
achieves only 31\% accuracy rate while Method~II achieves 69\%. In both cases, predictions 
are distributed broadly across multiple levels rather than concentrated near the true level, 
indicating persistent classification difficulty. This likely stems from limited Level~5 
representation in our dataset. The performance gap between the methods at Level~5 can be 
attributed to their fundamental difference in approach: Method~I relies on retrieving 
similar examples from the dataset, but with only 7\% of responses at Level~5, similarity 
search rarely surfaces relevant examples, leaving the model without meaningful context for 
classification. Method~II, by contrast, learns  representations through 
training, enabling it to capture Level~5 patterns  from sparse supervision. 

\section{Discussion and Conclusion}
This work demonstrates that Large Language Models can effectively infer teachers' Van Hiele reasoning levels when guided by structured pedagogical information. Our central hypothesis was that integrating explicit skills information improves Van Hiele classification. We tested this with two distinct approaches: Retrieval-Augmented Generation and Multi-Task Learning. In both cases, skills-aware variants significantly outperformed baselines without skills information across multiple evaluation metrics. We hypothesize that the skills dictionary helps models identify diagnostic patterns that distinguish Van Hiele levels. These include the use of logical language like "if...then" or "therefore," or linking properties across shape families. By defining reasoning patterns characteristic of each level, the skills help models attend to features that differentiate between levels.

Our research has practical implications for mathematics education research and professional development. By automating Van Hiele assessment - traditionally constrained by manual expert evaluation - our approach enables researchers to study geometric reasoning development at scale across large teacher cohorts. Automated assessment could also enable adaptive professional development systems that dynamically adjust content based on teachers' current Van Hiele levels and skill profiles. 
Additionally, extending the proposed models to output demonstrated skills alongside Van Hiele levels would create fine-grained diagnostic profiles for detecting strengths and gaps, enabling targeted interventions that address specific reasoning weaknesses.

Several limitations should be noted. First, the dataset of 226 responses is relatively small, particularly for higher Van Hiele levels, which may limit model generalizability. Second, our research was conducted with a fixed set of 59 questions, and models may not generalize to new problems; extending this approach to arbitrary geometry questions remains an important challenge for future work. Third, in contrast to Van Hiele levels, skills were identified through expert consensus rather than independent annotation; while our experiments validate their utility for Van Hiele classification, independent annotation with inter-rater reliability would strengthen confidence in skill labels. 

Beyond geometric reasoning, this work demonstrates a potentially generalizable approach: decomposing hierarchical learning frameworks into explicit, fine-grained skills that guide automated assessment. While we validated this methodology for Van Hiele levels, the principle may extend to other structured frameworks such as Bloom's Taxonomy or subject-specific reasoning models. Such approaches bridge AI capabilities with pedagogical theory by ensuring models assess learning using the same constructs educators use. Validating this skills-based methodology  remains important future work.

\begin{credits}
\subsubsection{\ackname}
This study was funded in part by Israeli Ministry of Science and Technology grant number 7774.
\subsubsection{\discintname}
The authors have no competing interests to declare that are relevant to the content of this article.
\end{credits}

\bibliographystyle{splncs04}
\bibliography{mybibliography}
\end{document}